\documentclass[letterpaper,journal]{IEEEtran}
\usepackage[T1]{fontenc}
\usepackage{amsmath,amssymb}
\usepackage{graphicx,booktabs,array,tabularx,cite}
\usepackage{tikz}
\usepackage{colortbl}
\definecolor{mvablue}{HTML}{2D91E6}
\definecolor{evalmuted}{HTML}{4E5964}
\definecolor{revisionred}{HTML}{000000} % Submission text: render revisions in black.
\DeclareRobustCommand{\rev}[1]{{\color{revisionred}#1}}
\definecolor{revisionblue}{HTML}{000000} % Clean submission: all revision text is black.
\DeclareRobustCommand{\newrev}[1]{{\color{revisionblue}#1}}
\newcommand{\revisedbib}[1]{\expandafter\def\csname revisedbib@#1\endcsname{}}
\revisedbib{malkov2018hnsw}
\revisedbib{douze2024faiss}
\revisedbib{secondhandsongs}
\revisedbib{daotwenty_midigpt}
\revisedbib{retkowski2025fmd}
\revisedbib{huang2020pop}
\revisedbib{raffel2016learning}
\revisedbib{choi2025deduplication}
\revisedbib{wu2023clamp}
\revisedbib{lee2022deduplicating}
\revisedbib{smith1981identification}
\revisedbib{gotoh1982improved}
\revisedbib{platt2000probabilities}
\revisedbib{ma2024symbolic}
\revisedbib{ji2024emomusictv}
\revisedbib{tian2025xmusic}
\revisedbib{wu2025clamp3}
\revisedbib{serra2008chroma}
\revisedbib{du2023bytecover3}
\revisedbib{foscarin2020asap}
\revisedbib{zhang2022atepp}
\revisedbib{wang2020pop909}
\revisedbib{long2025pdmx}
\revisedbib{lee2025gigamidi}
\revisedbib{goebl2017vienna}
\revisedbib{kim2025pianovam}
\revisedbib{pampalk2005artistfilter}
\revisedbib{sturm2012gtzan}
\revisedbib{sturm2013gtzanevaluation}
\revisedbib{rodriguezalgarra2019confounding}
\revisedbib{borovik2025symupe}
\usetikzlibrary{arrows.meta}
\usepackage{algorithm,algpseudocode}
\usepackage{epstopdf-base}
\usepackage[hidelinks]{hyperref}
\hypersetup{pdftitle={Beyond File-Level Splits: Rethinking Symbolic Music Evaluation via Multi-View Alignment},pdfsubject={Symbolic music evaluation and work-level isolation}}
\newcommand{\D}{\mathcal{D}}
\newcommand{\W}{\mathcal{W}}
\newcommand{\ind}{\mathbf{1}}

\title{Beyond File-Level Splits: Rethinking Symbolic Music Evaluation via Multi-View Alignment}

\author{Kun~Bai, Yinghan~Wang, Canran~Xiao, Jiachen~Luo, Huy~Phan, Yong~Dai, Jiaqi~Zhang, Yunwei~Li, Lin~Wang, Joshua~Reiss, and Björn~W. Schuller,~\IEEEmembership{Fellow,~IEEE}%
\thanks{Kun Bai is with the School of Intelligence Science and Technology, Beijing University of Civil Engineering and Architecture, Beijing 102616, China (e-mail: kunbai05@gmail.com).}%
\thanks{Yinghan Wang is with the College of Science, Beijing Forestry University, Beijing 100083, China (e-mail: yinghan@bjfu.edu.cn).}%
\thanks{Canran Xiao is with Sun Yat-sen University, Shenzhen 518107, China (e-mail: xiaocr3@mail.sysu.edu.cn).}%
\thanks{Jiachen Luo is with the School of Computation, Information and Technology, Technical University of Munich, Munich 80333, Germany, and also with the Centre for Digital Music, School of Electronic Engineering and Computer Science, Queen Mary University of London, London E1 4NS, UK (e-mail: jiachen.luo@qmul.ac.uk).}%
\thanks{Huy Phan is with the German Research Center for Artificial Intelligence (DFKI), Luebeck 23562, Germany (e-mail: huy.phan@dfki.de).}%
\thanks{Yong Dai is with the X-Humanoid, Beijing 101111, China (e-mail: daiyongya@outlook.com).}%
\thanks{Jiaqi Zhang is with the School of Computer Science and Telecommunication Engineering, Jiangsu University, Zhenjiang 212013, China (e-mail: 3230602065@stmail.ujs.edu.cn).}%
\thanks{Yunwei Li is with the University of Washington, Seattle, WA 98195, USA (e-mail: yunweili@uw.edu).}%
\thanks{Lin Wang and Joshua Reiss are with the Centre for Digital Music, School of Electronic Engineering and Computer Science, Queen Mary University of London, London E1 4NS, UK (e-mail: \{lin.wang,joshua.reiss\}@qmul.ac.uk).}%
\thanks{Björn W. Schuller is with the Chair of Health Informatics, Technical University of Munich, Munich 81675, Germany, and also with the Group on Language, Audio and Music, Imperial College London, London SW7 2AZ, UK (e-mail: schuller@tum.de).}%
}

\begin{document}
\maketitle
\begin{abstract}
\newrev{Symbolic music generation often continues MIDI passages by predicting subsequent notes or events.}
\rev{Held-out prediction loss can overestimate generalization to unseen compositions when training contains another performance or arrangement of a test work.}
\newrev{We introduce Multi-View Alignment-Aware Same-Work Identification, combining complementary musical representations with temporal alignment of shared passages to remove test-work versions from training while preserving unrelated material and fixing evaluation files.}
\newrev{In a controlled cohort of 100 manually verified pairs selected for high embedding similarity, replacing 3.29\% of training windows with test-work versions reduces mean per-group NLL by up to 30.16\% relative to matched unrelated content.}
\rev{Across two datasets and two generators, our method brings test losses closest to a reference built from independent work labels. Compared with removing the same number of files at random, it leaves fewer test works exposed and retains more eligible content, including about 99\% of eligible tokens on the piano-performance dataset.}
\rev{A separate study on that dataset matches training tokens and updates: no test works remain exposed under the dataset labels, and mean relative test loss is 2.15\% higher than an independently selected reference.}
\rev{These findings support keeping all versions of each composition in one split when evaluating generalization to unseen works.}
The project is available at \url{https://kunbai05-nb.github.io/}.\end{abstract}

\section{Introduction}
\label{sec:introduction}

\IEEEPARstart{S}{ymbolic} music generation models event sequences for composition, continuation, and accompaniment~\cite{huang2019music}, increasingly guided by user preferences and text~\cite{ma2024symbolic}. Coherent, controllable generation requires rhythmic and long-range dependencies~\cite{huang2020pop} and expressive control, such as emotion~\cite{ji2024emomusictv}. \rev{Evaluating generalization to unseen works must account for advantages gained from training exposure to test compositions.}

\begin{figure}[!tb]
    \centering
    \includegraphics[width=\linewidth]{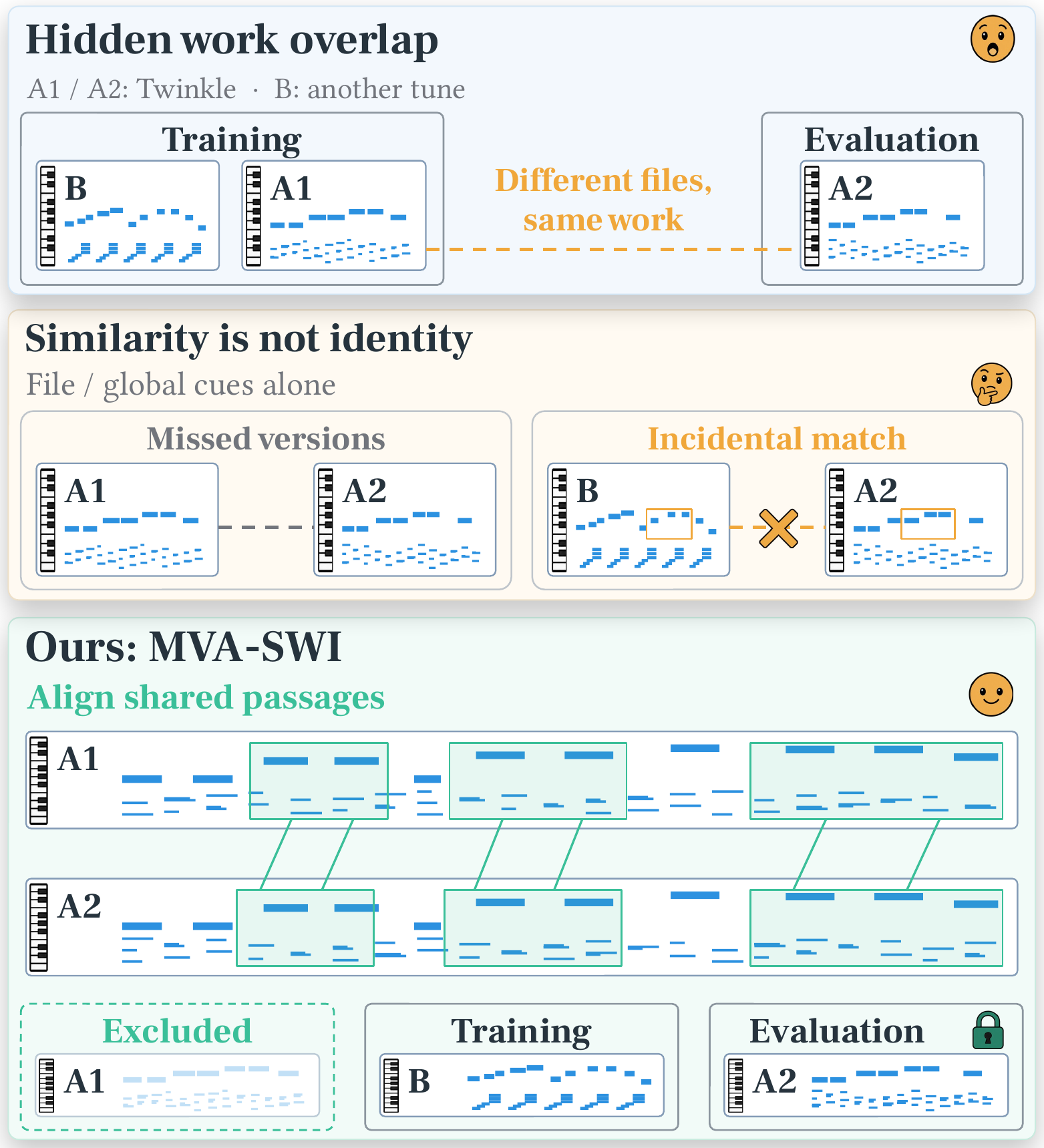}
    \caption{Illustrative example of work-level isolation. A1 and A2 are different performances of \emph{Twinkle, Twinkle, Little Star}, while B is another tune. MVA-SWI aligns shared passages to exclude A1 from training while retaining B and keeping the evaluation file A2 fixed.}
    \label{fig:teaser}
\end{figure}

Disjoint MIDI files can contain different performances, arrangements, or edits of the same composition~\cite{raffel2016learning}. A test file may therefore be absent from training while another version of its work is present, lowering held-out loss through exposure (Fig.~\ref{fig:teaser}). \rev{Sturm documented duplicate excerpts and alternate versions in GTZAN~\cite{sturm2012gtzan}; genre studies also showed that artist overlap can inflate scores~\cite{pampalk2005artistfilter,rodriguezalgarra2019confounding}.}

\rev{MAESTRO and SyMuPe's PERiScoPe keep all performances of a composition in one split~\cite{hawthorne2019maestro,borovik2025symupe}.} XMusic's XMIDI pipeline combines file hashes and chroma similarity for deduplication~\cite{tian2025xmusic}, while Aria-MIDI identifies repeated works from composition metadata~\cite{bradshaw2025aria}. \rev{Choi et al. develop complementary content-based MIDI deduplication methods, release evaluation code and filtered LMD lists, and identify cross-split versions as an evaluation threat~\cite{choi2025deduplication}.} Yet work identity remains difficult to recover from incomplete metadata and musically diverse versions. Shared motifs or harmony can also link distinct works, so isolation must distinguish exposure-relevant relations from legitimate resemblance.

Evaluation also requires understanding how version exposure changes generation losses across model families and how closely automatic cleaning approximates unseen-work evaluation. Missed relations leave test works exposed; false relations remove admissible content. Loss changes after cleaning therefore reflect both exposure and training-pool composition, requiring joint measurement of relation accuracy, retained content, and model losses.

We introduce Multi-View Alignment-Aware Same-Work Identification (MVA-SWI), combining complementary representations with temporal passage correspondences to distinguish versions from incidental similarity. \rev{With evaluation files fixed, we measure exposure effects and compare automatically isolated training with a reference built from independent work labels.}

Our contributions are threefold:
\begin{itemize}
    \item \textbf{Evaluation sensitivity.} We quantify model-dependent exposure effects under controlled training (Section~\ref{sec:controlled_exposure}). \newrev{In 100 manually verified, high-similarity pairs, replacing 3.29\% of training windows with test-work versions yields a mean per-group NLL reduction of up to 30.16\% (95\% CI: 27.35--33.02\%) relative to matched unrelated content.}
    \item \textbf{Content-based isolation.} MVA-SWI recovers version relations without reliable target-corpus work identifiers. We also audit detected cross-split exposure in seven MIDI collections.
    \item \textbf{Downstream validation.} We assess residual exposure, clean-content retention, and loss deviation from an independently labeled reference. For two generators on ATEPP and LMD-clean, MVA-SWI has the smallest mean absolute per-work relative NLL deviation among evaluated nonreference conditions. \rev{A budget-matched ATEPP study adds a reference selected independently of MVA-SWI; a separate study reports listener ratings and automatic exposure diagnostics.}
\end{itemize}

\section{Related Work}
\label{sec:related_work}

\subsection{Data Duplication and Evaluation Reliability}
Deduplication affects training-corpus composition and evaluation validity. \rev{On GTZAN, repeated recordings affected genre classifiers unevenly, and artist filtering changed their rankings~\cite{sturm2013gtzanevaluation}.} In language modeling, Lee et al.~\cite{lee2022deduplicating} compare exact substring matching with approximate document matching using $n$-gram overlap, linking repetition to memorization and train--validation overlap to optimistic evaluation. The relevant duplication granularity thus depends on the intended generalization.

File hashes and normalized event fingerprints capture exact or encoding-level repetition, whereas performances and arrangements can preserve work identity despite note changes~\cite{choi2025deduplication}. Aria-MIDI~\cite{bradshaw2025aria} addresses compositional duplication using composer and composition tags, including opus and piece numbers. Their availability and consistency limit metadata-based curation. We study the evaluation consequences of content-based isolation through residual exposure, clean-token retention, and loss deviation from an independently labeled reference. Fixed evaluation content and matched random removal help separate exposure and selection effects from training-pool size.

\subsection{Symbolic Music Retrieval and Deduplication}
Learned representations make symbolic music searchable beyond literal event matching. CLaMP~\cite{wu2023clamp} combines bar-level symbolic encoding with language--music contrastive learning, placing musical scores and descriptions in a shared space. CLaMP~3~\cite{wu2025clamp3} extends this direction across symbolic music, audio, and multilingual text. These representations retrieve files with shared musical attributes. Identifying versions of the same work further requires distinguishing persistent musical material from incidental resemblance.

\newrev{Choi et al.~\cite{choi2025deduplication} introduce CAugBERT, combine its predictions with CLaMP, and group duplicates by connected components. They already consider versions of a song and cross-split leakage, releasing evaluation code and filtered LMD lists. Building on this work-level scope, MVA-SWI adds ordered passage evidence to verification and evaluates partition isolation with held-out files fixed. We jointly measure residual exposure, eligible-token retention, and generation-loss deviation from independent-label references, and quantify the effect of supplying test-work versions through controlled training.}

\subsection{Music Version Matching and Local Alignment}
Serr\`a et al.~\cite{serra2008chroma} recover version matches using chroma similarity and local alignment, with gap penalties accommodating structural variation. Raffel~\cite{raffel2016learning} combines embedding-based pruning with dynamic time warping for audio--MIDI matching. ByteCover3~\cite{du2023bytecover3} uses local audio embeddings, approximate retrieval, and local-feature reranking for short-query identification, reducing interference from unrelated sections. Its MaxMean score averages segment-wise maximum similarities without enforcing temporal order.

Dataset isolation must control missed versions that leave exposure and false links that remove admissible content. MVA-SWI verifies pairs using order-preserving segment paths, calibrates decisions, and groups accepted relations by connected components. Order is enforced within each path; distinct regions may appear in different global orders. Rather than keeping one representative per group, partition isolation removes training files linked to fixed validation and test sets. Residual exposure and unnecessary removal measure the resulting errors.

\section{Motivation and Problem Formulation}
\label{sec:problem}

\begin{table*}[!t]
\centering
\caption{Controlled exposure with fixed test content. U/R/S denote Unrelated/Repeated-unrelated/Same. \newrev{The last column averages within-group U-to-S reductions~\eqref{eq:exposure_effect}, with paired 95\% work-group bootstrap CIs conditional on three trained seeds and the selected high-similarity cohort.} Dataset labels define R; two reviewers verify each nominated pair. -- denotes unavailable conditions. Comparisons are within each model and objective.}
\label{tab:controlled_exposure}
\small
\setlength{\tabcolsep}{4pt}
\begin{tabular*}{\textwidth}{@{\extracolsep{\fill}}llrrrrrrr@{}}
\toprule
& & \multicolumn{4}{c}{Mean loss} & \multicolumn{2}{c}{Control contrasts} & Relative effect \\
\cmidrule(lr){3-6}\cmidrule(lr){7-8}\cmidrule(l){9-9}
Model & Objective & Clean & Unrelated (U) & Repeated (R) & Same (S) & R$-$U & R$-$S & \newrev{Reduction [95\% CI] (\%)} \\
\midrule
Transformer-S & NLL & 0.816 & 0.813 & 0.817 & 0.657 & +0.004 & +0.160 & 20.30 \newrev{[17.62, 23.13]} \\
Transformer-M & NLL & 0.812 & 0.800 & 0.810 & 0.638 & +0.010 & +0.172 & 21.46 \newrev{[18.65, 24.44]} \\
Transformer-L & NLL & 0.798 & 0.791 & 0.816 & 0.589 & +0.025 & +0.227 & 27.09 \newrev{[23.75, 30.58]} \\
MIDI-GPT & NLL & 0.736 & 0.732 & 0.747 & 0.526 & +0.015 & +0.221 & 30.16 \newrev{[27.35, 33.02]} \\
LSTM & NLL & 1.220 & 1.210 & 1.237 & 1.153 & +0.027 & +0.084 & 4.74 \newrev{[3.69, 5.88]} \\
TCN-384 & NLL & 1.235 & 1.234 & 1.237 & 1.184 & +0.003 & +0.053 & 4.04 \newrev{[3.40, 4.75]} \\
Conditional VAE & NLL & 0.860 & 0.848 & -- & 0.839 & -- & -- & 1.04 \newrev{[0.44, 1.67]} \\
% Source MSE values (Clean/U/S): 0.01525828 / 0.01526124 / 0.01525690; display uses three decimal places throughout the loss columns.
Latent Diffusion & MSE & \newrev{0.015} & \newrev{0.015} & -- & \newrev{0.015} & -- & -- & 0.01 \newrev{[${-}0.07$, 0.10]} \\
 
\bottomrule
\end{tabular*}
\end{table*}

\subsection{From File Disjointness to Work Independence}
Let $\D_{\rm tr}$, $\D_{\rm va}$, and $\D_{\rm te}$ denote training, validation, and test MIDI files. We evaluate generation on \emph{previously unseen works}: the compositions underlying performances and versions. Let $g(x)$ be file $x$'s reference work identity, when available, and $\W(\D)=\{g(x):x\in\D\}$. File disjointness requires $\D_{\rm tr}\cap\D_{\rm te}=\varnothing$; work independence requires $\W(\D_{\rm tr})\cap\W(\D_{\rm te})=\varnothing$.
A test work is \emph{exposed} if any version occurs in training, even in a different file. Shared identity may instead be intentional when evaluating new performances of known works. \rev{Reference labels use corpus work identifiers and repertoire metadata independently of detector outputs and downstream losses (labeling rules and audit scope: Appendix~\ref{app:work_labels}).}

Let $\ell_w^{(c)}$ be condition $c$'s loss on fixed test content from work $w$. NLL is averaged over valid scored tokens per file, equally over files per work, then over paired training seeds. Other objectives keep scoring units and evaluation content fixed. Controlled exposure supplies another test-work version ($c={\rm same}$) or unrelated content ($c={\rm unrelated}$) at identical replacement positions and token budgets. For positive baseline losses, the mean relative exposure advantage is
\begin{equation}
 R=\frac{100}{|\W_{\rm te}|}
 \sum_{w\in\W_{\rm te}}
 \frac{\ell_w^{({\rm unrelated})}-\ell_w^{({\rm same})}}
      {\ell_w^{({\rm unrelated})}},
 \label{eq:exposure_effect}
\end{equation}
where $\W_{\rm te}=\W(\D_{\rm te})$. \rev{Each controlled-cohort unit is a confirmed same-work test--donor pair; the average pertains to this selected cohort.} Relative changes are averaged within a fixed model and loss definition.

\subsection{Controlled Evidence of Exposure Effects}
\label{sec:controlled_exposure}
In a fixed LMD cohort, 1,264 of 38,374 training windows (3.29\%) occupy fixed replacement positions. Clean retains their content, Unrelated substitutes unrelated content, and Same supplies another version of each test work. Unrelated and Same match positions and total input tokens. Evaluation fixes 100 groups, each with one test file and one donor, three corresponding training seeds, and 20,000 optimization steps. \newrev{A frozen CAugBERT $\cap$ CLaMP graph (both thresholds 0.99) nominates high-similarity pairs. Two reviewers confirm all 100 as same-work, including six full-work--excerpt pairs (Appendix~\ref{app:work_labels}). This selected cohort demonstrates exposure sensitivity; it does not estimate the average effect across all works or version types.} A label-based Repeated-unrelated control matches repetition counts, length, and note density to isolate repetition effects.

Exposure effects vary by model (Table~\ref{tab:controlled_exposure}). Relative to Unrelated, mean within-group NLL reductions are 20.30\%, 21.46\%, and 27.09\% for Transformer-S/-M/-L, 30.16\% for MIDI-GPT, and 4.74\%/4.04\% for LSTM/TCN. \newrev{The paired 95\% CI for MIDI-GPT is [27.35, 33.02]\%; diffusion's MSE reduction is 0.01\% [${-}0.07$, 0.10]. VAE decoder NLL and diffusion noise MSE probe different objectives, so these values do not rank model families by general exposure sensitivity. Additional MIDI-GPT runs at 5,000 and 10,000 updates yield 4.39\% and 14.44\% reductions (Appendix~\ref{app:training_duration}).}

\newrev{All six autoregressive configurations have lower loss under Same than either unrelated control. For example, MIDI-GPT's NLL is 0.747 under Repeated-unrelated versus 0.526 under Same, retaining an exposure advantage after matching repetition.}

\subsection{Relation Recovery and Isolation Criteria}
When $g$ is unavailable, we estimate same-work relations $r(x_i,x_j)=\ind[g(x_i)=g(x_j)]$ from MIDI content. These define groups $\hat g$, evaluated against $g$ where available; for file set $S$, let $\hat g(S)=\{\hat g(x):x\in S\}$. We assess isolation by retained content and subsequent model losses.

For retained training set $\D_{\rm tr}^{\rm iso}$, independent labels measure residual exposure:
\begin{equation}
 E_{\rm work}=
 \frac{|\W(\D_{\rm tr}^{\rm iso})\cap\W_{\rm te}|}
      {|\W_{\rm te}|}.
 \label{eq:exposure}
\end{equation}
To measure unnecessary removal, define the admissible training subset $\D_{\rm safe}=\{x\in\D_{\rm tr}:g(x)\notin\W(\D_{\rm va}\cup\D_{\rm te})\}$. If $t(x)$ counts usable training tokens under a fixed tokenizer and preprocessing protocol and $\sum_{x\in\D_{\rm safe}}t(x)>0$, clean-token retention is
\begin{equation}
 K_{\rm token}=\frac{\sum_{x\in\D_{\rm tr}^{\rm iso}\cap\D_{\rm safe}}t(x)}
                         {\sum_{x\in\D_{\rm safe}}t(x)}.
 \label{eq:retention}
\end{equation}
Missed relations can leave exposure; false relations can reduce clean-token retention.

For downstream evaluation, a reference model is trained on independently label-isolated data. For each test work with positive reference loss, define
\begin{equation}
\delta_w=\frac{\ell_w^{({\rm iso})}-\ell_w^{({\rm ref})}}{\ell_w^{({\rm ref})}}.
\label{eq:relative_deviation}
\end{equation}
For a nonempty test-work set, the signed deviation $B_{\rm signed}$ is the work-level mean of $100\delta_w$, and the absolute deviation is
\begin{equation}
 B_{\rm abs}=\frac{100}{|\W_{\rm te}|}\sum_{w\in\W_{\rm te}}
 |\delta_w|.
 \label{eq:deviation}
\end{equation}
Absolute deviations prevent cancellation across works. Agreement depends on training composition and volume; matched random removal helps separate relation-guided selection from data reduction.

\section{Multi-View Alignment for Work-Level Isolation}
\label{sec:method}

\begin{figure*}[t]
\centering
\includegraphics[width=\textwidth]{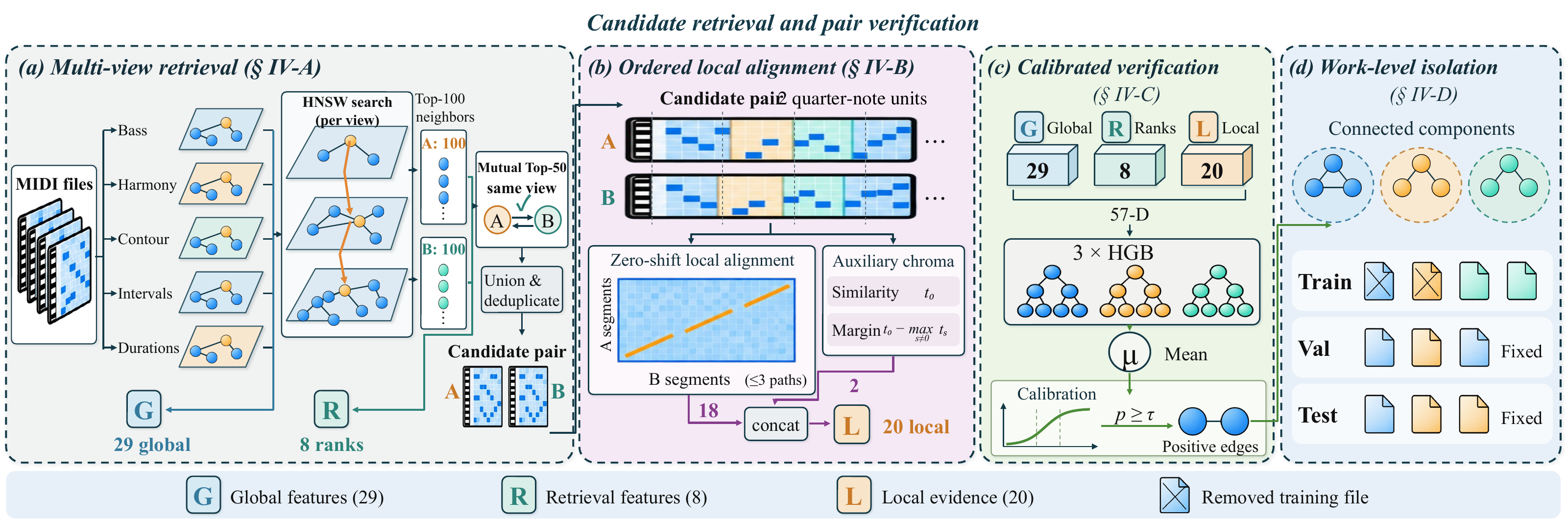}
\caption{Overview of the MVA-SWI workflow. (a) HNSW retrieves up to 100 neighbors per view for rank features; pairs that are mutual Top-50 neighbors within the same view are united across the five views. (b) Zero-shift local alignment supplies 18 path features, complemented by two chroma auxiliary features. (c) The 57-dimensional evidence vector is scored by a classifier ensemble and sigmoid calibration. (d) Connected components of threshold-positive edges guide training-data removal while validation and test sets remain fixed.}
\label{fig:pipeline}
\end{figure*}

MVA-SWI isolates training content from fixed validation and test sets by recovering version relations. Bass patterns, harmony, or melodic motifs may persist despite other changes. We retrieve candidates through complementary views, verify ordered local correspondences, and group calibrated pairwise decisions (Fig.~\ref{fig:pipeline}). Inference uses MIDI content from all three partitions; work labels support development and evaluation only. Algorithm~\ref{alg:overview} summarizes frozen-verifier inference; Appendix~\ref{app:representations} specifies parsing and feature fallbacks.

\begin{algorithm}[!b]
\caption{Core MVA-SWI inference and work isolation}
\label{alg:overview}
\small
\begin{algorithmic}[1]
\Require Fixed partitions $(\D_{\rm tr},\D_{\rm va},\D_{\rm te})$
\Require Frozen $f_1,f_2,f_3$, calibration $(a,b_0)$, threshold $\tau$
\Ensure Inferred groups $\hat g$ and retained training set $\D_{\rm tr}^{\rm iso}$
\State $\D\gets\D_{\rm tr}\cup\D_{\rm va}\cup\D_{\rm te}$
\State Build five retrieval views; fit IDF on $\D$ (Appendix~\ref{app:representations})
\State Build one HNSW index per view; retrieve Top-100 nonself neighbors for globally valid files
\State Form $\mathcal C$ from mutual Top-50 pairs within any single view by~\eqref{eq:candidates}; store each pair once as $i<j$
\For{each $\{i,j\}\in\mathcal C$}
  \State Compute 29 global and eight retrieval features
  \State Compute 18 path features from zero-shift alignment (Appendix~\ref{app:alignment}) and two auxiliary chroma features
  \State Assemble $z_{ij}\in\mathbb R^{57}$ in Appendix~\ref{app:features} order
  \State $\hat p_{ij}\gets\sigma\!\left(a\,\frac{1}{3}\sum_{b=1}^{3}f_b(z_{ij})+b_0\right)$
\EndFor
\State $\mathcal E^+\gets\{\{i,j\}\in\mathcal C:\hat p_{ij}\geq\tau\}$
\State Find connected components of $(\D,\mathcal E^+)$ and isolate training files by~\eqref{eq:isolation}
\State \Return $\hat g,\D_{\rm tr}^{\rm iso}$
\end{algorithmic}
\end{algorithm}

\subsection{Multi-View Candidate Retrieval}
\label{sec:retrieval}
Five views represent each file: bass-interval $n$-grams, transposition-canonicalized harmonic sequences, upper-contour motifs, and interval and duration histograms. Token views encode recurring patterns using nonnegative hashing, sublinear term frequency, and inverse document frequency (IDF); histograms capture pitch and timing distributions. All views are $\ell_2$ normalized. IDF adapts to the current retrieval corpus without work labels, while the verifier stays fixed (parsing, quantization, and weighting: Appendix~\ref{app:representations}).

\rev{Faiss~\cite{douze2024faiss} builds a hierarchical navigable small world (HNSW) index~\cite{malkov2018hnsw} per view and queries each valid file.} Token vectors are projected and renormalized; histograms are indexed directly. Inner-product search approximates cosine-neighbor retrieval. Unprojected vectors supply global verification features; Appendix~\ref{app:representations} specifies projection and index parameters.

Let $N_v^k(i)$ denote the first $k$ nonself neighbors returned by HNSW for file $x_i$ in view $v$. We retain a pair only when both files retrieve one another within the same view:
\begin{equation}
\mathcal C=\bigcup_{v\in\mathcal V}
\bigl\{\{i,j\}:i<j,\ j\in N_v^{50}(i),\ i\in N_v^{50}(j)\bigr\}.
\label{eq:candidates}
\end{equation}
Both retrieval directions must occur in one view; any view suffices. Each query retains up to 100 nonself neighbors: the first 50 determine candidacy, and all supply verification ranks. Per-view mutual sets are united and deduplicated, storing each pair once by fixed input-file index. Training, calibration, and inference share this rule.

\subsection{Order-Preserving Local Alignment}
\label{sec:alignment}
To test whether shared material forms coherent musical regions, we compare sequences of short segments. \newrev{After shifting the first note onset to zero, we divide MIDI ticks by the header's pulses per quarter note (PPQ) and form segments of two MIDI quarter-note units ($2\times\mathrm{PPQ}$ ticks). This uses neither the tempo map nor inferred performance beats.} At most the first 512 segments are retained. Each segment records upper and bass pitches in eight onset bins, an onset-count vector, a pitch-class profile, and a note count. A note contributes only to its onset segment, with pitch-class weight equal to its velocity multiplied by its duration clipped to \newrev{two quarter-note units}.

For segments $u$ and $v$, melody similarity $m_{uv}$ averages $\exp(-|p-q|/2.5)$ over corresponding bins containing upper-contour MIDI pitches $p$ and $q$. Bass similarity $b_{uv}$ is computed analogously; either similarity is zero when no bin is jointly valid. Harmony similarity $h_{uv}$ and rhythm similarity $r_{uv}$ are inner products of the $\ell_2$-normalized pitch-class and onset-count vectors, respectively. Given note counts $a_u$ and $a_v$, density similarity is $d_{uv}=\exp(-|\log(1+a_u)-\log(1+a_v)|)$. We combine them as
\begin{equation}
 s_{uv}=0.30m_{uv}+0.15b_{uv}+0.25h_{uv}
       +0.20r_{uv}+0.10d_{uv}.
\label{eq:segment_similarity}
\end{equation}
Local alignment uses the observed pitches at zero pitch shift. \newrev{Zero shift refers to pitch, not tempo. Alignment permits displaced passages and gaps but applies no explicit tempo normalization; rubato and tempo differences can still impair correspondence (Appendix~\ref{app:tempo_sensitivity}).} Global transposition-invariant representations and aggregate chroma evidence provide separate verification features.

We recover local correspondences with an affine-gap dynamic program~\cite{smith1981identification,gotoh1982improved}. Let $q_{uv}=2(s_{uv}-0.525)$, let $H_{uv}$ be the best local score ending at $(u,v)$, and let $E_{uv}$ and $F_{uv}$ be the two gap states. Starting from zero boundary states,
\begin{align}
E_{uv}&=\max\{0,H_{u-1,v}-\gamma_o,E_{u-1,v}-\gamma_e\},\nonumber\\
F_{uv}&=\max\{0,H_{u,v-1}-\gamma_o,F_{u,v-1}-\gamma_e\},\nonumber\\
H_{uv}&=\max\{0,H_{u-1,v-1}+q_{uv},E_{uv},F_{uv}\},
\label{eq:alignment}
\end{align}
where $\gamma_o=0.35$ and $\gamma_e=0.08$. Traceback from the maximum-scoring state yields a path of matched segment pairs. Zero restarts allow a match to begin inside either file, while gap states accommodate intervening unmatched segments.

We extract at most three paths. If the highest-scoring path contains fewer than four matched pairs, extraction terminates. Otherwise, its matched rows and columns are masked before the next search. Each path preserves temporal order, while distinct paths can occur in different global orders. For retained sequence lengths $n_A,n_B$ and matched-index sets $I_A,I_B$, the coverage values are $c_A=|I_A|/n_A$ and $c_B=|I_B|/n_B$. Their harmonic mean is $c_{\rm hm}=2c_Ac_B/(c_A+c_B)$, or zero when both coverages vanish. Coverage is measured over the retained prefixes. Appendix~\ref{app:alignment} gives the extraction procedure, including traceback ties and stopping rules.

\subsection{Calibrated Same-Work Verification}
\label{sec:verification}
\newrev{Each candidate receives a 57-dimensional vector $z_{ij}$: 29 global comparisons, eight retrieval statistics, and 20 local features. Global features compare musical patterns, distributions, and statistics; retrieval features summarize ranks and mutual support. The local block has 18 path-derived coordinates and two alignment-independent chroma features: zero-shift aggregate similarity and its margin over the best nonzero shift. Appendix~\ref{app:features} defines all coordinates and boundary cases.}

The target includes files from the same recording, but development uses one file per recording identifier, no repeated checksum, and disjoint training, calibration, and test works. Positive pairs share a work across recording identifiers; negatives have different works. The pairwise benchmark also excludes same-recording pairs and thus measures cross-recording identification. Unlabeled-corpus inference applies no recording-based exclusion (cohorts and scope: Appendix~\ref{app:learning}).

We retain all positive candidate pairs and independently sample negatives for three class-balanced histogram-based gradient-boosting classifiers. Let $f_b(z_{ij})\in[0,1]$ be classifier $b$'s predicted positive-class probability. Their arithmetic mean is calibrated directly by a sigmoid~\cite{platt2000probabilities}:
\begin{equation}
\bar f_{ij}=\frac{1}{3}\sum_{b=1}^{3}f_b(z_{ij}),\qquad
\hat p_{ij}=\sigma(a\bar f_{ij}+b_0),
\label{eq:calibration}
\end{equation}
where $\sigma(t)=(1+e^{-t})^{-1}$. The full calibration candidate pool is used both to fit $(a,b_0)$ and to choose the threshold $\tau$. Threshold selection maximizes calibration F1 subject to precision of at least $0.97$ and at least 20 predicted-positive pairs. The classifiers, calibrator, and threshold are fitted for the specified feature representation and HNSW candidate rule, and are then frozen for evaluation on each target corpus. End-to-end evaluation counts eligible same-work pairs omitted by retrieval as false negatives. Sampling, optimization, and threshold tie rules are detailed in Appendix~\ref{app:learning}.

\subsection{Relation-Guided Work Isolation}
\label{sec:isolation}
An undirected graph over all files has threshold-positive edges $\mathcal E^+=\{\{i,j\}\in\mathcal C:\hat p_{ij}\geq\tau\}$. Connected components define work groups, obtained by uniting singleton groups along every positive edge, independently of processing order (Appendix~\ref{app:grouping}).

Let $\hat g$ denote the connected-component assignment. Keeping validation and test files fixed, we retain
\begin{equation}
\D_{\rm tr}^{\rm iso}=\{x\in\D_{\rm tr}:\hat g(x)\notin
\hat g(\D_{\rm va}\cup\D_{\rm te})\}.
\label{eq:isolation}
\end{equation}
Every accepted relation has its endpoints in the same inferred group, so no threshold-positive edge connects the retained training set to a held-out file. False-positive bridges can merge unrelated works, while missing edges can fragment one work. Residual work exposure and clean-content retention measure these errors under independent labels using~\eqref{eq:exposure} and~\eqref{eq:retention}. For unlabeled corpora, overlap under $\hat g$ is reported as detected group exposure. Files excluded by global parsing remain unresolved singleton groups.

For $N$ files and $V$ views, mutual Top-$k$ retrieval generates at most $VkN/2$ unordered candidate pairs before cross-view deduplication. HNSW avoids an exhaustive all-pairs similarity scan; its search cost and recall depend on the index and search budgets. Each alignment pass costs $O(n_An_B)$ time and memory, with at most three passes. Connected components can be obtained by union--find in $O((N+|\mathcal E^+|)\alpha(N))$ time, where $\alpha$ is the inverse Ackermann function. Appendix~\ref{app:variants} distinguishes the core detector from the separate configuration used for the seven-dataset audit.

\section{Experiments}
\label{sec:experiments}

We evaluate relation recovery, ablations, and detected overlap in seven MIDI collections. Independently labeled studies measure residual exposure, clean-token retention, and NLL agreement with a label-isolated reference; \rev{listening assesses perceived musicality, while automatic diagnostics measure similarity to held-out content.}

\subsection{Experimental Setup}
\label{sec:experimental_setup}
Fixed-threshold comparisons use ATEPP~\cite{zhang2022atepp}, ASAP~\cite{foscarin2020asap}, LMD-clean~\cite{choi2025deduplication}, and MAESTRO~\cite{hawthorne2019maestro}. Calibration also covers \rev{SecondHandSongs (SHS)~\cite{secondhandsongs}}, Vienna $4\times22$~\cite{goebl2017vienna}, and PianoVAM~\cite{kim2025pianovam}, but excludes MAESTRO. Each comparison fixes its cohorts. We use frozen v1.9 (Section~\ref{sec:method}), fitted on work-disjoint LMD partitions: five-view HNSW retrieval, same-view mutual Top-50 candidates, zero-shift alignment, and three classifiers with 57 features. The threshold is $0.597596$; positive-edge components define isolation groups. Appendix~\ref{app:learning} details fitting and configuration; Appendix~\ref{app:variants} describes the separate corpus auditor.

Baselines are file SHA-256, normalized event fingerprints, CAugBERT, CLaMP, and CAugBERT $\cup$ CLaMP~\cite{choi2025deduplication,wu2023clamp}, accepting either model's positive prediction at 0.93/0.87. Isolation also tests their intersection at 0.99 for both, \newrev{a calibrated union, and metadata matching where available.} Pairwise precision (P), recall (R), and F1 exclude same-recording pairs and count unretrieved positives as false negatives. Feature and retrieval results are unweighted macro averages over LMD-clean, ATEPP, and ASAP. Isolation uses~\eqref{eq:exposure}--\eqref{eq:deviation}; Appendix~\ref{app:experiment_protocols} defines intervals and variation across splits and classifier subsets.

\begin{table*}[!t]
\centering
\caption{Pairwise identification at fixed operating points. P, R, and F1 are percentages; FN counts missed reference-positive pairs, including unretrieved pairs. CAugBERT and CLaMP use thresholds 0.93 and 0.87, respectively; $\cup$ denotes their prediction union. MVA-SWI uses 0.597596. A dash denotes undefined precision when no positive pair is predicted.}
\label{tab:pairwise_fixed}
\small
\setlength{\tabcolsep}{2.4pt}
\begin{tabularx}{\textwidth}{>{\raggedright\arraybackslash}X*{16}{r}}
\toprule
& \multicolumn{4}{c}{ATEPP} & \multicolumn{4}{c}{ASAP} & \multicolumn{4}{c}{LMD-clean} & \multicolumn{4}{c}{MAESTRO} \\
\cmidrule(lr){2-5}\cmidrule(lr){6-9}\cmidrule(lr){10-13}\cmidrule(l){14-17}
Method & P & R & F1 & FN & P & R & F1 & FN & P & R & F1 & FN & P & R & F1 & FN \\
\midrule
SHA-256 & -- & 0.00 & 0.00 & 3339 & 100.00 & 0.02 & 0.04 & 4987 & -- & 0.00 & 0.00 & 2037 & -- & 0.00 & 0.00 & 880 \\
Event fingerprint & -- & 0.00 & 0.00 & 3339 & 100.00 & 0.02 & 0.04 & 4987 & 96.72 & 20.27 & 33.52 & 1624 & -- & 0.00 & 0.00 & 880 \\
CAugBERT & 12.69 & 31.84 & 18.14 & 2276 & 35.67 & 50.36 & 41.76 & 2476 & 94.09 & 28.92 & 44.24 & 1448 & 62.45 & 38.18 & 47.39 & 544 \\
CLaMP & 62.71 & 69.54 & 65.95 & 1017 & 86.64 & 30.03 & 44.60 & 3490 & 94.64 & 26.02 & 40.82 & 1507 & 88.45 & 53.98 & 67.04 & 405 \\
CAugBERT $\cup$ CLaMP & 22.86 & 76.61 & 35.21 & 781 & 40.34 & 64.43 & 49.61 & 1774 & 94.13 & 33.04 & 48.91 & 1364 & 70.48 & 70.00 & 70.24 & 264 \\
\rowcolor{mvablue!7} \textbf{MVA-SWI} & 98.08 & 97.72 & 97.90 & 76 & 99.84 & 99.62 & 99.73 & 19 & 96.18 & 67.99 & 79.67 & 652 & 100.00 & 100.00 & 100.00 & 0 \\
 
\bottomrule
\end{tabularx}
\end{table*}

\begin{figure}[!tb]
\centering
\includegraphics[width=\columnwidth]{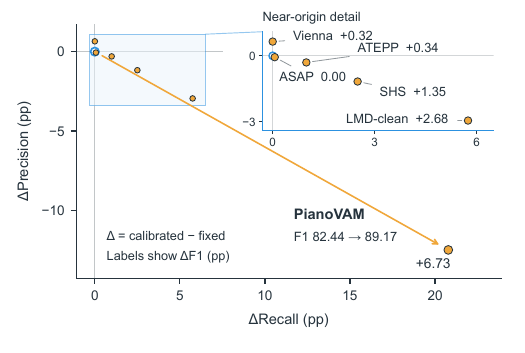}
\caption{Calibration gains and precision--recall tradeoffs. Coordinates show calibrated-minus-fixed recall and precision in percentage points (pp); numeric labels give the corresponding F1 change. Vienna denotes Vienna $4\times22$. \newrev{Table~\ref{tab:calibration_key} gives the calibrated 95\% F1 confidence intervals.}}
\label{fig:threshold_calibration}
\end{figure}

\begin{table}[!tb]
\centering
\caption{Verifier ablations: macro percentages over LMD-clean, ATEPP, and ASAP. $\textcolor{black}{\checkmark}$/$\textcolor{black}{\times}$: used/omitted; Clf.: classifier; D: input dimension. $\dagger$: mutual Top-100 support in at least two bass/harmony/motif views; candidates remain same-view mutual Top-50. Feature-removal variants are retrained and recalibrated. AP uses candidates; recall includes unretrieved positives.}
\label{tab:feature_ablation}
\small
\setlength{\tabcolsep}{2pt}
\begin{tabularx}{\columnwidth}{*{4}{>{\centering\arraybackslash}X}rrrrr}
\toprule
\multicolumn{4}{c}{Verifier components} & & \multicolumn{4}{c}{Macro performance (\%)} \\
\cmidrule(lr){1-4}\cmidrule(l){6-9}
Global & Retr. & Local & Clf. & D & P & R & F1 & AP \\
\midrule
\rowcolor{mvablue!7} $\textcolor{black}{\checkmark}$ & $\textcolor{black}{\checkmark}$ & $\textcolor{black}{\checkmark}$ & $\textcolor{black}{\checkmark}$ & 57 & 98.03 & 88.45 & 92.43 & 95.95 \\
$\textcolor{black}{\times}$ & $\textcolor{black}{\checkmark}$ & $\textcolor{black}{\checkmark}$ & $\textcolor{black}{\checkmark}$ & 28 & 98.37 & 85.14 & 90.51 & 94.99 \\
$\textcolor{black}{\checkmark}$ & $\textcolor{black}{\times}$ & $\textcolor{black}{\checkmark}$ & $\textcolor{black}{\checkmark}$ & 49 & 97.68 & 87.37 & 91.49 & 95.27 \\
$\textcolor{black}{\checkmark}$ & $\textcolor{black}{\checkmark}$ & $\textcolor{black}{\times}$ & $\textcolor{black}{\checkmark}$ & 37 & 97.99 & 82.07 & 88.32 & 94.53 \\
\midrule $\textcolor{black}{\times}$ & $\textcolor{black}{\checkmark}^{\dagger}$ & $\textcolor{black}{\times}$ & $\textcolor{black}{\times}$ & -- & 43.29 & 88.45 & 57.30 & 75.59 \\
 
\bottomrule
\end{tabularx}
\end{table}

\subsection{Same-Work Identification and Calibration}
\label{sec:identification_results}

Exposure control requires recovering cross-version relations while limiting incorrect links that can remove admissible content. At fixed thresholds, MVA-SWI achieves F1 scores of 97.90\%, 99.73\%, 79.67\%, and 100.00\% on ATEPP, ASAP, LMD-clean, and MAESTRO, respectively (Table~\ref{tab:pairwise_fixed}). Exact matching recovers almost no cross-recording relations outside a subset of LMD-clean pairs. Embedding similarities recover more positives, with dataset-dependent precision--recall balance. On ATEPP, CAugBERT $\cup$ CLaMP improves recall over either constituent but reaches only 22.86\% precision; MVA-SWI attains 98.08\% precision at 97.72\% recall. On LMD-clean, MVA-SWI misses 652 of 2,037 reference-positive pairs despite 96.18\% precision, motivating direct measurement of residual exposure in the isolation experiments.

We compare independent target-dataset calibration with direct transfer of the fixed LMD threshold (Fig.~\ref{fig:threshold_calibration}). MVA-SWI has higher F1 point estimates than calibrated CAugBERT $\cup$ CLaMP on all six cohorts (\newrev{Table~\ref{tab:calibration_key}}). Calibration increases its F1 from 79.67\% to 82.35\% on LMD-clean and from 82.44\% to 89.17\% on PianoVAM, while ASAP remains at 99.73\%. On PianoVAM, recall rises from 70.13\% to 90.91\% as precision falls from 100.00\% to 87.50\%. \newrev{SHS reaches only 72.99\% calibrated F1 (95\% CI: 51.28--85.71\%), contrasting with 100\% fixed-threshold F1 on MAESTRO. Together with LMD-clean's missed positives, this limits claims of uniform robustness across heterogeneous sources.}

\begin{figure*}[!t]
\centering
\includegraphics[width=\textwidth]{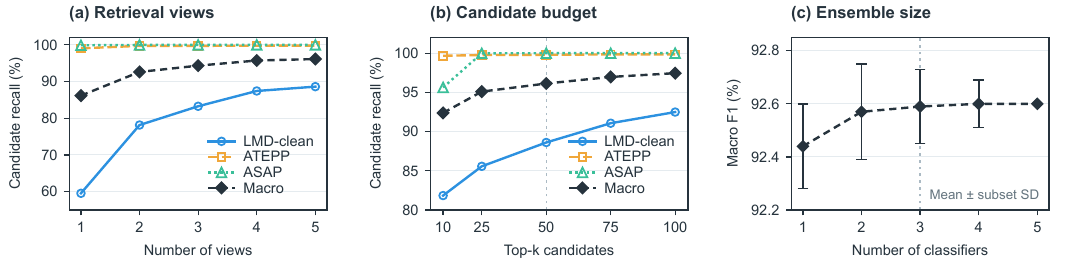}
\caption{Retrieval coverage by dataset and macro ensemble F1 over three datasets. (a) View subsets are independently selected on MAESTRO-validation at each size. (b) Candidate recall as the neighbor budget Top-$k$ increases. (c) Error bars are population standard deviations over all subsets of five trained classifiers, with calibration refitted for each subset. Zero spread at five classifiers reflects a single combination.}
\label{fig:retrieval_sensitivity}
\end{figure*}

\subsection{Contributions of Retrieval, Alignment, and Verification}
\label{sec:ablation_results}

Candidate recall rises from 86.16\% with one selected view to 96.13\% with five; LMD-clean gains most, from 59.55\% to 88.61\% (Fig.~\ref{fig:retrieval_sensitivity}a). This supports complementary views for heterogeneous data. Top-100 adds only 1.32 points of macro recall over the deployed Top-50 budget (Fig.~\ref{fig:retrieval_sensitivity}b).

With candidates fixed, removing the 20 local features produces the largest feature-block F1 drop, 92.43\% to 88.32\%, and lowers recall from 88.45\% to 82.07\% (Table~\ref{tab:feature_ablation}). Removing global or retrieval features reduces F1 by 1.92 or 0.94 points. A mutual-neighbor rule replacing learned verification gives 43.29\% precision and 57.30\% F1. \rev{The local block includes two alignment-independent chroma features, so this ablation does not isolate temporal ordering. Shift and prefix tests use the full vector (\newrev{\href{https://kunbai05-nb.github.io/}{Extended Results}}).}

Increasing the ensemble from three to five classifiers changes mean macro F1 from 92.59\% to 92.60\%, averaged over all combinations at each size (Fig.~\ref{fig:retrieval_sensitivity}c). The 92.43\% full-model result (Table~\ref{tab:feature_ablation}) and isolation use the core system's three fixed deployment seeds.
\rev{At 512 segments, the Top-3 shift variant gains 3.63 percentage points of recall and loses 2.17 points of precision on LMD-clean. F1 decreases by 0.18 and 0.15 points on ATEPP and ASAP, respectively. Extending the zero-shift prefix to 768 segments changes F1 by at most 0.02 points.}
\newrev{For pairs differing by over 20\% in annotated median beat interval, ASAP recall is 98.32\%; ATEPP recall is 94.50\% under a note-span duration proxy. Appendix~\ref{app:tempo_sensitivity} reports strata and fixed-verifier parameter sensitivity.}

\begin{figure}[!tb]
\centering
\includegraphics[width=\columnwidth]{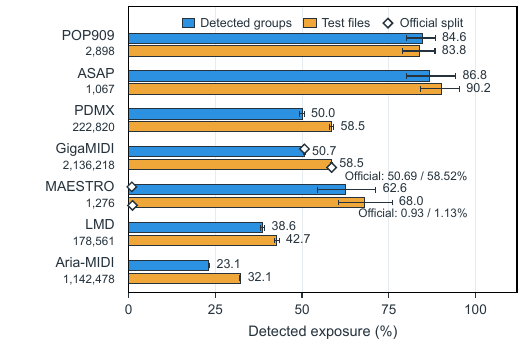}
\caption{Detected exposure across seven MIDI collections. Bars show mean group/file exposure over 200 random 80/10/10 file splits; whiskers show 2.5th--97.5th empirical percentiles. Diamonds and group/file labels indicate official splits; counts below the names give the number of input files. \rev{Intervals quantify split variability conditional on the inferred groups.}}
\label{fig:corpus_exposure}
\end{figure}

\begin{table*}[!t]
\centering
\caption{\newrev{Training-pool isolation with default and calibrated baselines, averaged over paired random splits.} $E$ is residual work exposure (\%, lower is better); $K$ is clean-token retention (\%, higher is better). The $\cup$ and $\cap$ baselines use union and intersection, respectively; threshold pairs list CAugBERT first and CLaMP second. \newrev{-- denotes an unreported metadata baseline.}}
\label{tab:cleaning_fixed}
\small
\setlength{\tabcolsep}{3pt}
% Use horizontal-mode numeric columns so color changes cannot shift row baselines.
% All values have two decimal places, so right alignment also aligns decimal points.
\newcommand{\isolationnumwidth}{\dimexpr(\textwidth-46pt-20\tabcolsep)/10\relax}
\begin{tabular}{@{}w{l}{46pt}*{2}{>{\columncolor{mvablue!7}}w{r}{\isolationnumwidth}}*{4}{w{r}{\isolationnumwidth}}*{4}{>{\color{revisionblue}}w{r}{\isolationnumwidth}}@{}}
\toprule
& \multicolumn{2}{>{\columncolor{mvablue!7}}c}{MVA-SWI} & \multicolumn{2}{c}{CAugBERT $\cup$ CLaMP} & \multicolumn{2}{c}{CAugBERT $\cap$ CLaMP} & \multicolumn{2}{c}{\newrev{CAugBERT $\cup$ CLaMP}} & \multicolumn{2}{c}{\newrev{Metadata matching}} \\
& \multicolumn{2}{>{\columncolor{mvablue!7}}c}{} & \multicolumn{2}{c}{Default} & \multicolumn{2}{c}{Intersection} & \multicolumn{2}{c}{\newrev{Calibrated}} & \multicolumn{2}{c}{} \\
& \multicolumn{2}{>{\columncolor{mvablue!7}}c}{(0.597596)} & \multicolumn{2}{c}{(0.93/0.87)} & \multicolumn{2}{c}{(0.99/0.99)} & \multicolumn{2}{c}{\newrev{(0.964/0.876)}} & \multicolumn{2}{c}{} \\
\cmidrule(lr){2-3}\cmidrule(lr){4-5}\cmidrule(lr){6-7}\cmidrule(lr){8-9}\cmidrule(l){10-11}
Dataset & $E\downarrow$ & $K\uparrow$ & $E\downarrow$ & $K\uparrow$ & $E\downarrow$ & $K\uparrow$ & $E\downarrow$ & $K\uparrow$ & $E\downarrow$ & $K\uparrow$ \\
\midrule
ATEPP & 1.24 & 99.59 & 18.60 & 31.35 & \newrev{99.47} & 100.00 & 44.74 & 75.57 & 22.56 & 95.56 \\
ASAP & 1.61 & 98.82 & 34.81 & 42.66 & 90.69 & 100.00 & 72.54 & 89.18 & -- & -- \\
LMD-clean & 17.75 & 99.19 & 35.66 & 99.35 & \newrev{49.74} & \newrev{99.90} & 36.49 & 99.43 & -- & -- \\
MAESTRO & 0.00 & 100.00 & 20.46 & 66.56 & 67.63 & 100.00 & 41.94 & 84.51 & 2.01 & 100.00 \\
 
\bottomrule
\end{tabular}
\end{table*}

\begin{figure}[!tb]
\centering
\includegraphics[width=\columnwidth]{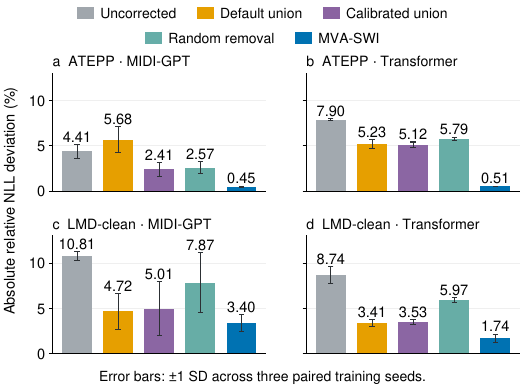}
\caption{\newrev{Absolute per-work relative NLL deviation $B_{\rm abs}$ from the independent-label reference (lower is closer), including the calibrated union baseline. Bars show means and whiskers $\pm1$ SD across three paired training seeds. Training-pool details and the distinction from Fig.~\ref{fig:matched_reference} are in Appendix~\ref{app:downstream_details}.}}
\label{fig:downstream_deviation}
\end{figure}

\begin{figure*}[!t]
\centering
\includegraphics[width=\textwidth]{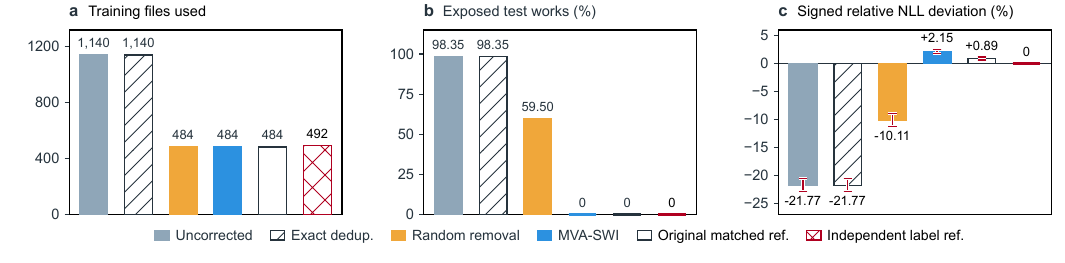}
\caption{\rev{Budget-controlled ATEPP isolation: (a) training files, (b) test works exposed under ATEPP labels (\%), and (c) mean per-work relative NLL deviation from the independent-label reference (\%), with paired 95\% intervals. Negative deviations indicate lower loss. Every condition uses 20,000 updates, 11,316 training windows, and 5,884,369 scored tokens.}}
\label{fig:matched_reference}
\end{figure*}

\subsection{Detected Exposure Across MIDI Collections}
\label{sec:corpus_audit}
We audit seven frozen corpus snapshots: POP909~\cite{wang2020pop909}, ASAP, PDMX~\cite{long2025pdmx}, GigaMIDI~\cite{lee2025gigamidi}, MAESTRO, LMD~\cite{raffel2016learning}, and Aria-MIDI~\cite{bradshaw2025aria}, totaling 3,685,318 input files. Approximate retrieval, a preliminary gate, and an independently trained 49-feature verifier remain fixed across corpora (Appendix~\ref{app:variants}). This configuration differs from the 57-feature benchmark detector and measures overlap between predicted groups.

A test group is exposed if it contains a training file; test-file exposure weights predicted groups by their test-file counts. Under random file splits, mean detected-group exposure ranges from 23.13\% for Aria-MIDI to 86.77\% for ASAP, and mean test-file exposure from 32.11\% to 90.23\% (Fig.~\ref{fig:corpus_exposure}). GigaMIDI's official split yields 50.69\% group exposure and 58.52\% file exposure, close to the random-split means of 50.71\% and 58.52\%. MAESTRO's official split yields only 0.93\%/1.13\%, versus 62.56\%/67.98\% under random splits. Thus, file partitions can retain substantial detected overlap, which split construction can reduce. \rev{In the three-cohort labeled benchmark, the 49-feature Fast configuration attains pair precision of 91.70--99.70\% and recall of 62.89--99.66\%, with the largest F1 gap to v1.9 on LMD-clean (74.61\% versus 79.67\%). In the deployed audit of four unlabeled corpora, reviewers confirm 91--95 of 100 randomly sampled accepted edges and 15--17 of 20 targeted bridges per corpus. Nine of 20 selected large components contain confirmed members from more than one work. Appendix~\ref{app:corpus_audit} reports these checks, parsing failures, and end-to-end audit costs.}

\subsection{Isolation Quality and Downstream Evaluation}
\label{sec:isolation_results}
We assess removal through independently labeled exposure and retention, and loss agreement with a label-isolated reference.

\noindent\textbf{Residual exposure and content retention.} Only the training pool is cleaned under paired random 80/10/10 splits, with 20 partitions for the three primary sources (Appendix~\ref{app:cleaning_protocol}). At the fixed operating point, MVA-SWI retains at least 98.82\% of clean tokens in all four datasets, with residual exposure of 1.24\% on ATEPP, 1.61\% on ASAP, 17.75\% on LMD-clean, and 0.00\% on MAESTRO (Table~\ref{tab:cleaning_fixed}). \newrev{The calibrated union improves clean-token retention over the default union on every dataset but leaves more exposure, reaching 44.74\% on ATEPP and 72.54\% on ASAP. Metadata matching leaves 22.56\%/2.01\% exposure on ATEPP/MAESTRO. The high-threshold intersection preserves more content while missing more test-work versions. MVA-SWI has the lowest exposure throughout, with slightly lower LMD-clean retention than the embedding baselines.}

A leave-one-source study tests threshold transfer. A 1\% development exposure target yields 12.95\% exposure and 92.63\% clean-token retention on the held-out LMD-clean source; some stricter targets admit no feasible threshold (Appendix~\ref{app:threshold_transfer}). Transferred thresholds therefore require assessment of target-cohort exposure and retention.

\noindent\textbf{Agreement with the label-isolated reference.} Using fixed cohorts distinct from the repeated-split averages above, we train MIDI-GPT and a Transformer on ATEPP and LMD-clean under \newrev{six conditions: Uncorrected, matched random removal, MVA-SWI isolation, default and calibrated CAugBERT $\cup$ CLaMP isolation, and an independent-label reference.} Each uses three training seeds and fixed evaluation content within each dataset. Random removal matches MVA-SWI's retained-file counts of 486 on ATEPP and 1,023 on LMD-clean, separating content selection from file count. Other conditions retain different amounts of data (Appendix~\ref{app:downstream_details}). Agreement is measured by the absolute per-work relative NLL deviation $B_{\rm abs}$ in~\eqref{eq:deviation}; smaller values indicate closer agreement.

MVA-SWI has the smallest \newrev{mean $B_{\rm abs}$ among the five nonreference conditions} in all four dataset--model settings (Fig.~\ref{fig:downstream_deviation}). On ATEPP, deviations are 0.45\% for MIDI-GPT and 0.51\% for Transformer, compared with 2.57\% and 5.79\% under matched random removal. MVA-SWI exposes none of the 121 test works and retains 98.91\% of clean tokens; random removal exposes 72 works and retains 63.97\%. On LMD-clean, the corresponding MVA-SWI deviations are 3.40\% and 1.74\%, versus 7.87\% and 5.97\% under random removal. It retains 99.67\% of clean tokens while exposing 25 of 142 test works. \newrev{The calibrated union yields 2.41\%/5.12\% on ATEPP and 5.01\%/3.53\% on LMD-clean (MIDI-GPT/Transformer). These mean comparisons include seed SDs but do not establish significant differences between conditions.}

\begin{figure}[!tb]
\centering
\includegraphics[width=\columnwidth]{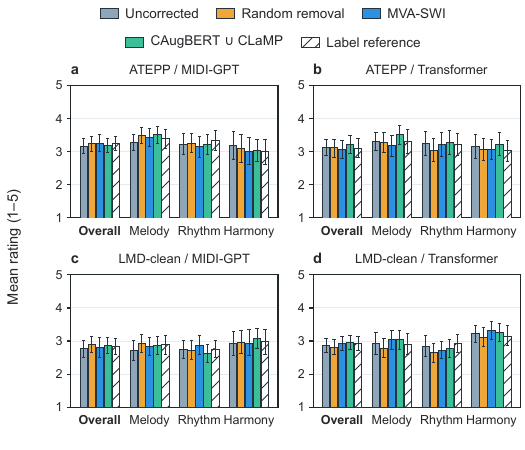}
\caption{Blinded ratings (1--5; higher is better). Bars show means with 95\% CIs for overall musicality (primary), melodic coherence, rhythmic naturalness, and harmonic coherence.}
\label{fig:subjective_quality}
\end{figure}

\noindent\textbf{Controlling the training budget.} \rev{A separate six-condition ATEPP study matches training updates, windows, and scored tokens (Fig.~\ref{fig:matched_reference} and Appendix~\ref{app:matched_reference}). MVA-SWI leaves 0 of 121 test works exposed under ATEPP labels and yields a mean relative NLL deviation of $+2.15\%$ from the independently selected reference (paired 95\% CI $[+1.85\%,+2.48\%]$).}
\newrev{Figures~\ref{fig:downstream_deviation} and~\ref{fig:matched_reference} use different training pools, references, and budget protocols. Their absolute and signed deviations are separate estimands, not alternative summaries of the same runs.}

\subsection{Perceptual Quality and Generation Diagnostics}
\label{sec:listening_results}
\rev{Using MIDI-GPT and Transformer on ATEPP and LMD-clean, we describe listener ratings for the five downstream conditions and separately report automatic continuation diagnostics (Fig.~\ref{fig:subjective_quality} and \newrev{Table~\ref{tab:passage_key}}).} \newrev{These five conditions exclude the newly added calibrated union.} This study uses ten training seeds, distinct from the three used for isolation, and listening prompts from the same test pools as the cleaning experiments. Appendix~\ref{app:subjective_protocol} details stimulus assignment and statistical analysis.

\noindent\textbf{Blinded listening.} Eighty retained listeners, half with and half without a musical background, provide 4,800 formal audio evaluations. \rev{Listeners heard only the generated continuations, which we rendered from MIDI with FluidSynth at 120 BPM and a 16 kHz sample rate. They gave informed consent before rating the excerpts.} Each anonymous eight-second excerpt receives five-point ratings for overall musicality, melodic coherence, rhythmic naturalness, and harmonic coherence. Figure~\ref{fig:subjective_quality} reports condition means and 95\% confidence intervals; paired contrasts use overall musicality as the primary outcome. \rev{The MVA-SWI-minus-Uncorrected differences on ATEPP are $+0.09$ ($[-0.09,+0.27]$) for MIDI-GPT and $-0.05$ ($[-0.25,+0.12]$) for Transformer. On LMD-clean they are $+0.05$ ($[-0.14,+0.23]$) and $+0.06$ ($[-0.12,+0.23]$), respectively. These 95\% intervals account for listener and work variation conditional on the ten trained seeds. The contrasts are descriptive and leave a range of positive and negative effects compatible with the data.}

\noindent\textbf{Automatic diagnostics.} Automatic evaluation uses fixed sets of 120 ATEPP works and 119 LMD-clean reference groups. Melody and rhythm similarity measure agreement with true test continuations; \rev{FMD-CLaMP1 applies Fr\'echet Music Distance~\cite{retkowski2025fmd} to CLaMP embeddings~\cite{wu2023clamp} of generated and shared natural continuations.} Across the four dataset--model settings, MVA-SWI reduces long-passage match rates to $2.46$--$4.06\%$, versus $21.90$--$23.89\%$ for Uncorrected and $10.48$--$11.69\%$ for matched random removal; label-isolated rates are $2.41$--$3.94\%$ (\newrev{Table~\ref{tab:passage_key}}). Its mean melody and rhythm similarities are within $0.44$ and $0.34$ percentage points of the reference, respectively, and its FMD is within $0.18$ (\newrev{\href{https://kunbai05-nb.github.io/}{Extended Results}}). \rev{Uncorrected yields higher continuation similarity and lower FMD in every setting. Because the comparison uses the held-out continuations as the target, these measurements describe agreement with that target and can reflect work exposure. Listener ratings provide a separate perceptual outcome.}

\section{Conclusion}
\label{sec:conclusion}

\newrev{MVA-SWI addresses a key challenge in symbolic music evaluation: different performances or arrangements of the same composition can appear in both training and test sets. It combines complementary musical representations with temporal alignment and calibrated verification to identify these versions and remove them from training while keeping the evaluation files fixed.}

\newrev{Our controlled study of 100 manually verified, high-similarity pairs shows why this matters: exposure to another version can lower prediction loss, with effects that vary across models. When training budgets are matched on ATEPP, MVA-SWI removes all exposure identified by the dataset labels, and mean relative test loss is $2.15\%$ higher than an independently selected reference. Isolation remains imperfect on LMD-clean, where some exposure persists. Generated continuations also contain fewer long-passage matches in all four evaluated settings. The listening results, however, remain inconclusive about differences in musical quality and do not establish equivalence.}

\newrev{Missed versions can leave overlap undetected, while incorrect links can remove valid training material. Corpus-wide estimates also depend on the auditor's coverage, and detected overlap alone does not establish copyright infringement. These limitations reinforce the need to report remaining overlap, retained training content, and musical quality alongside prediction loss. Together, our findings support keeping versions of each composition in one split when evaluating generalization to unseen works.}

\begingroup
\newcommand*{\bibdoi}[1]{doi: \href{https://doi.org/#1}{\nolinkurl{#1}}}
\let\originalbibitem\bibitem
\renewcommand{\bibitem}[1]{%
  \originalbibitem{#1}%
  \ifcsname revisedbib@#1\endcsname\color{revisionred}\else\color{black}\fi}
\bibliographystyle{IEEEtran}
\bibliography{references,notes/experiment_sources/dataset_references,references_related_work_new,references_intro_tmm_new}
\endgroup
\clearpage
\appendices
% Condensed supplement. Full prior sources: appendix/archive/20261031_before_four_pages/.
\section{Representations and Pair Features}
\label{app:representations}

\subsection{Parsing, Tokenization, and Retrieval}
\newrev{Times and durations are measured in MIDI quarter-note units by dividing ticks by PPQ, without using the tempo map or estimating performance beats.} Channel-10 notes are excluded when nonpercussion notes exist; otherwise they are retained. Same-channel/pitch note-ons are closed in first-in, first-out order, with minimum duration one tick; unmatched note-ons receive \newrev{0.25 quarter-note units}. Files without notes are invalid. Rounding uses nearest integers with ties to even. Global onset bins are $\operatorname{round}(4t)$, retaining leading silence. Their highest/lowest pitches define upper/bass contours; consecutive intervals are clipped to $[-36,36]$. Occupied-bin gaps $\Delta t$ yield rhythm values $\min\{192,\max\{0,\operatorname{round}(12\Delta t)\}\}$.

Upper/bass tokens are interval $n$-grams, $n=2,3,4,5$; rhythm uses $n=2,3,4$. Motifs contain three consecutive upper-interval/IOI pairs. For $C>0$ available starts, sample $0,d,2d,\ldots<C$ with $d=\max\{1,\lceil C/K\rceil\}$ and $K=2048$. Harmony uses onset-only pitch-class masks in bins $\operatorname{round}(2t)$, retaining empty bins between the first and last occupied bins, capped at 2048 bins from the first. Length-2, 4, and 8 mask sequences are canonicalized to their lexicographically smallest form under 12 common pitch-class rotations, with $K=512$ starts per length.

Each token family uses nonnegative \texttt{FeatureHasher} counts in 8192 bins, weights $(1+\log c)\{1+\log[(N+1)/(\mathrm{df}+1)]\}$, and $\ell_2$ normalization. IDF is fitted within the retrieval corpus; $N$ includes invalid zero rows. Interval histograms have 75 bins indexed by $\delta+37$. Duration/IOI histograms have 16 bins indexed by
\[
\operatorname{clip}\bigl(\operatorname{round}[2\log_2\max(t,1/32)]+8,0,15\bigr).
\]
Duration counts use all notes and IOI counts use successive occupied bins. Full-file chroma weights pitch class $p\bmod12$ by $\max(d,0.125)\max(v,1)$ for duration $d$ and velocity $v$. Histograms are $\ell_2$ normalized; zero vectors remain zero.

Token retrieval vectors use sparse random projection to $\min\{256,\max(8,N_{\rm valid}-1)\}$ dimensions, followed by FP32 storage and renormalization. Seeds are $20260911+v$, with views ordered bass, harmony, motif, interval, duration; histogram views are indexed directly. Each FAISS \texttt{IndexHNSWFlat} uses inner products, $M=32$, \texttt{efConstruction}$=160$, and \texttt{efSearch}$=200$, with fixed insertion order and one thread. Query up to 101 entries, remove self, and retain at most 100 neighbors. Fewer than two valid files give no candidates. Equation~\eqref{eq:candidates} applies same-view mutual Top-50 filtering. Global features use unprojected vectors.

\subsection{The 57 Coordinates and Boundary Cases}
\label{app:features}
For each of upper contour, bass, rhythm, harmony, and motif, the first 15 coordinates are cosine, Jaccard, and containment, in that order. With occupied-bin sets $S_A,S_B$ and $I=|S_A\cap S_B|$, the latter two are
\[
J=\frac{I}{\max(|S_A\cup S_B|,1)},\quad
K=\frac{I}{\max(\min(|S_A|,|S_B|),1)}.
\]
Next come interval, duration, and IOI cosines, then maximum-shift global chroma cosine over 12 rotations. The remaining ten global coordinates compare, in order, notes, occupied onset bins, final offset, tracks, pitch minimum/maximum/mean/population standard deviation, distinct programs, and occupied \newrev{half-quarter-note bins}. Pitch statistics use absolute differences; other scalars use $\min(a,b)/\max(a,b,10^{-6})$. Track counts follow the percussion fallback above.

Next append $(R_v,M_v)$ for bass, harmony, and motif:
\[
R_v=\frac{1}{\min(r_{AB}^v,r_{BA}^v)},\quad
M_v=\ind[\max(r_{AB}^v,r_{BA}^v)\leq100].
\]
Ranks beyond 100 or absent neighbors are 101. These six coordinates precede the local block; $\sum_vM_v$ and $\max_vR_v$ finish the vector, giving eight retrieval coordinates overall.

For retained paths $\ell=1,\ldots,P$, let $Q_\ell,m_\ell,U_\ell,V_\ell$ denote score, match count, and matched-index sets, and $M=\sum_\ell m_\ell$. Define
\[
\begin{gathered}
c_A=\frac{|\cup_\ell U_\ell|}{n_A},\qquad
c_B=\frac{|\cup_\ell V_\ell|}{n_B},\\
C=\frac{2c_Ac_B}{\max(c_A+c_B,10^{-8})}.
\end{gathered}
\]
The combined span is $T_\ell=\max U_\ell-\min U_\ell+\max V_\ell-\min V_\ell+2$. The match-weighted gap fraction is
\[
G=\sum_\ell\frac{m_\ell}{M}\frac{\max(0,T_\ell-2m_\ell)}{T_\ell}.
\]
The 14 path coordinates are, in order: $S=\max_\ell(Q_\ell/m_\ell)$, $\sum_\ell Q_\ell/M$, $\max_\ell m_\ell$, $M$, $P$, short/long coverage, $C$, $G$, and the five matched-pair mean similarities $(a_m,a_b,a_r,a_h,a_d)$ from~\eqref{eq:segment_similarity}. Coverage uses retained sequence lengths, assigning $A$ to the short entry on a tie.

For normalized segment chroma $h_{Au}$, let $\bar h_A$ be the normalized sum $\sum_u h_{Au}$, and define $t_s=\bar h_A^{\mathsf T}\operatorname{roll}(\bar h_B,s)$. Append $t_0$ and $t_0-\max_{s\ne0}t_s$, followed by four interactions: $\min(a_m,a_b,a_r,a_h)$, $(a_m+a_b+a_r+a_h)/4$, $SC$, and $S(1-G)$. Thus the local block contains 18 path-derived and two independent chroma coordinates. With valid sequences but no accepted path, only the two chroma values survive; extraction failure zeros all 20 local coordinates. Zero vectors/empty sets give zero similarities. Features are FP32, replacing NaN/$+\infty$/$-\infty$ by 0/1/0. Canonical input order $A<B$ fixes alignment orientation; returned HNSW order fixes rank ties.

\section{Alignment, Learning, and Grouping}
\subsection{Local Sequence and Traceback}
\label{app:alignment}
After subtracting the earliest onset, retain $n=\min\{512,\max(1,\lceil T/2\rceil)\}$ \newrev{segments of two MIDI quarter-note units}, where $T$ is the final offset \newrev{in the same units}. A note at $t$ enters segment $u=\lfloor t/2\rfloor$ and bin $\min\{7,\lfloor4(t-2u)\rfloor\}$; discard $u\geq n$ and retain internal empty segments. Each bin records highest/lowest onset pitches and all onset counts. Chroma adds $\min(d,2)\max(v,1)$ at the note's pitch class, entirely within its onset segment. Segment chroma and onset-count vectors are normalized separately. Pitch similarities average jointly valid bins; no jointly valid bins give zero. Two empty segments have zero pitch, rhythm, and harmony similarity, but unit density similarity.

Apply~\eqref{eq:alignment} at zero pitch shift with zero boundary states. Scan rows then columns, retaining the first maximum. Positive-state ties prefer diagonal, $E$, then $F$; gap ties prefer opening over extension. Traceback stops at a zero restart or boundary and emits diagonal matches. After each accepted path, set its matched rows and columns in the reward matrix to $-10^6$; subsequent paths can cross them only through gaps. Stop after three paths, a nonpositive maximum, or a best path with fewer than four matches, even if a lower-scoring admissible path exists.

\subsection{Work-Disjoint Fitting and Calibration}
\label{app:learning}
\label{app:core_configuration}
The LMD development cohort has 16,788 checksum-unique files and 10,098 works, with one file per recording identifier. Hash each work identifier using the first eight hexadecimal SHA-256 digits modulo 100. Buckets $[0,70)$, $[70,85)$, and $[85,100)$ contain respectively 11,740/7,049, 2,501/1,500, and 2,547/1,549 files/works for fitting, calibration, and testing. Fitting uses five consecutive 14-bucket shards. IDF, indexing, and candidate selection run separately per shard; concatenate their feature rows for training. Calibration/test pools are independently constructed. Positives share a work; negatives have different works. Recording/checksum uniqueness is a development restriction, not an inference filter.

For seeds 20260911--20260913, retain all 8,634 positive candidates and sample up to 50 negatives per positive without replacement, yielding 440,334 examples per classifier. Shuffle and fit class-balanced \texttt{HistGradientBoostingClassifier} models with 260 iterations, learning rate 0.05, at most 15 leaves, minimum leaf size 80, $\ell_2$ strength 8, and no early stopping (scikit-learn 1.7.2; NumPy 1.26.4).

Fit logistic regression to the mean classifier probability on the full calibration pool (1,800 positives), using \texttt{lbfgs}, $C=1$, and at most 1,000 iterations. The fitted values are $a=9.548566767981713$ and $b_0=-9.098677121144501$; clip the sigmoid argument to $[-50,50]$. Evaluate distinct scores in descending order, accepting whole tie groups. Require precision $\geq0.97$ and at least 20 accepted pairs, then maximize (F1, recall, precision) lexicographically, breaking remaining ties by descending threshold. If infeasible, return no threshold. The frozen operating point is $\tau=0.5975961974276196$. Representation/candidate changes require refitting classifiers, calibration, and threshold.

\subsection{Isolation and the Separate Corpus Auditor}
\label{app:grouping}
\label{app:variants}
Initialize singleton groups and union every threshold-positive edge. Edge order does not affect connected components; there is no size cap, score ordering, or bridge-count constraint. Apply~\eqref{eq:isolation} with validation/test fixed. Invalid files remain unresolved singletons. False bridges can remove admissible content, while missed relations can leave exposure.

The seven-corpus auditor uses the same retrieval--alignment structure but requires mutual Top-20 support in at least two views, a 29-feature preliminary gate, and an independently trained 49-feature verifier at score threshold 0.9396. This configuration is frozen across corpora and is distinct from the calibrated 57-feature core detector. Its stricter retrieval/gating can omit core candidates; reported exposure is conditional on its inferred groups. \rev{The deployment replay processes edges in descending score order, requires multiple positive links for larger merges, and caps group size. The released deployment configuration records these grouping parameters.}

\section{\newrev{Experimental Protocols and Key Results}}
\label{app:experiment_protocols}

\subsection{Cohorts, Calibration, and Sensitivity}
\newrev{Pairwise cohorts contain files/eligible positives: SHS 306/80, ASAP 1,067/4,988, ATEPP 1,527/3,339, LMD-clean 2,547/2,037, Vienna $4\times22$ 88/924, and PianoVAM 94/77. Positives share a work but differ in recording ID; predictions use the same exclusion. Recall includes retrieval misses. Table~\ref{tab:calibration_key} reports independently calibrated F1. Its 95\% CIs use 2,000 work-group bootstrap replicates (seed 20260921), pooling query-level TP/FP/FN; each unordered pair contributes twice before resampling. LMD-clean is the work-disjoint test cohort in Appendix~\ref{app:learning}.}
\begin{table}[!tb]
\centering
\caption{\newrev{Calibrated F1 (\%) and work-bootstrap 95\% CIs. Union: CAugBERT $\cup$ CLaMP. Full calibration results: \href{https://kunbai05-nb.github.io/}{Extended Results on the project page}.}}
\label{tab:calibration_key}
\color{revisionblue}
\small
\setlength{\tabcolsep}{2.5pt}
\begin{tabular*}{\columnwidth}{@{\extracolsep{\fill}}lrrr@{}}
\toprule
Dataset & MVA-SWI & 95\% CI & Union \\
\midrule
SHS & 72.99 & [51.28, 85.71] & 39.68 \\
ASAP & 99.73 & [99.40, 99.93] & 60.40 \\
ATEPP & 98.24 & [97.15, 99.09] & 68.20 \\
LMD-clean & 82.35 & [78.77, 85.81] & 51.12 \\
Vienna $4\times22$ & 100.00 & [100.00, 100.00] & 90.92 \\
PianoVAM & 89.17 & [71.50, 95.09] & 57.33 \\
\bottomrule
\end{tabular*}
\end{table}

\newrev{Feature ablations retrain/recalibrate seeds 20260911--20260913 under the same precision constraint, fixing candidates; full/no-global/no-retrieval/no-local dimensions are 57/28/49/37. AP uses scored candidates. Sensitivity averages LMD-clean, ATEPP, and ASAP. At each view count, MAESTRO-validation selects the subset by most retrieved positives, then fewest candidates, then lexicographic view order; that subset transfers unchanged. Top-$k$ uses all views at $k=10,25,50,75,100$. Both require same-view reciprocal support and exclude same-recording pairs. Ensemble sizes 1--5 recalibrate all 5/10/10/5/1 subsets of seeds 20260911--20260915. Means/population SDs summarize overlapping subsets; the three-classifier mean differs from the fixed three-seed ablation.}

\subsection{\newrev{Alignment and Timing Checks}}
\label{app:alignment_robustness}
\label{app:tempo_sensitivity}
\newrev{Shift/prefix tests keep candidates and all 57 features fixed (\href{https://kunbai05-nb.github.io/}{Extended Results}). A separate constructed two-hop test retains 54/54 true and 0/174 false links with shared-position filtering, versus 29/54 and 103/174 after position shuffling. This is not part of frozen grouping and does not isolate ordered versus unordered matching with identical features. Late-only passages and phrase reordering remain unevaluated.}

\newrev{ASAP timing strata use the larger/smaller median annotated interbeat interval; ATEPP uses note-span duration, which also reflects content (\href{https://kunbai05-nb.github.io/}{Extended Results}). Annotations are diagnostic only; strata include retrieval misses and do not measure local rubato. With candidates, classifier, and threshold fixed, we scale both gap penalties by 0.5/1.5, test equal and melody-/harmony-emphasized weights, and use 1/2/4-quarter-note segments at fixed input coverage. Across ASAP, ATEPP, and LMD-clean, F1 changes by at most 0.6 percentage points and P/R by less than one point. Equal/default weights perform similarly; alternatives are not separately optimized.}

\subsection{\newrev{Independent Labels and Exposure Controls}}
\label{app:work_labels}
\label{app:cleaning_protocol}
\label{app:training_duration}
\newrev{Independent labels use ATEPP composition IDs, ASAP score directories, LMD-clean artist/song groups, MAESTRO composer/title, Vienna score IDs, PianoVAM work metadata, and SHS cover-work groups~\cite{zhang2022atepp,foscarin2020asap,choi2025deduplication,hawthorne2019maestro,goebl2017vienna,kim2025pianovam,secondhandsongs}. Performances, transcriptions, transpositions, arrangements, and unambiguous excerpts share a work label; movements, medleys, and quotations are flagged separately. Title/style/motif similarity alone is insufficient. Two reviewers initially confirmed 98 and 99 of the 100 selected pairs, marking the rest uncertain. Joint review confirmed all 100, including six work--excerpt pairs.}

\newrev{For the controlled study, Clean excludes the selected test/donor components and file identities. Unrelated/Same replace identical windows with matched tokens. Average paired-seed losses within each group before computing its relative reduction; groups receive equal weight. Table~\ref{tab:controlled_exposure}'s CIs use 10,000 group-bootstrap replicates with trained seeds fixed, excluding selection and new-training uncertainty. Separately retrained 5k/10k runs keep the 20k learning-rate schedule but use RTX 3090/4090 rather than RTX 5090, limiting attribution to update count alone (\href{https://kunbai05-nb.github.io/}{Extended Results}). Cleanup averages 20 paired 80/10/10 partitions for ATEPP, ASAP, and LMD-clean; only training files are removed. Cleanup timing excludes detection.}

\subsection{Corpus Audit and Threshold Transfer}
\label{app:corpus_audit}
\label{app:threshold_transfer}
\newrev{Table~\ref{tab:auditor_key} independently validates the 49-feature Fast auditor against the 57-feature core. The seven-corpus replay uses a separate checkpoint/grouping configuration. Across 200 random 80/10/10 splits, percentile intervals describe split variability conditional on inferred groups; official splits use manifest metadata. Missed relations can lower estimated exposure and false bridges can raise it.}
\begin{table}[!tb]
\centering
\caption{\newrev{Independent validation of the Fast auditor and v1.9 core. Pairwise P, R, and F1 are percentages; recall includes retrieval misses.}}
\label{tab:auditor_key}
\color{revisionblue}
\small
\setlength{\tabcolsep}{2.5pt}
\begin{tabular*}{\columnwidth}{@{\extracolsep{\fill}}llrrr@{}}
\toprule
Cohort & Model & P & R & F1 \\
\midrule
ATEPP & Fast & 98.37 & 97.36 & 97.86 \\
 & v1.9 & 98.08 & 97.72 & 97.90 \\
LMD-clean & Fast & 91.70 & 62.89 & 74.61 \\
 & v1.9 & 96.18 & 67.99 & 79.67 \\
ASAP & Fast & 99.70 & 99.66 & 99.68 \\
 & v1.9 & 99.84 & 99.62 & 99.73 \\
\bottomrule
\end{tabular*}
\end{table}

\begin{table}[!tb]
\centering
\caption{\newrev{Manual audit of unlabeled corpora. Edges/bridges report same-work/different-work/uncertain counts. Mixed components contain more than one confirmed work.}}
\label{tab:manual_audit_key}
\color{revisionblue}
\small
\setlength{\tabcolsep}{2.5pt}
\begin{tabular*}{\columnwidth}{@{\extracolsep{\fill}}lrrr@{}}
\toprule
Corpus & \shortstack[r]{Random\\edges} & \shortstack[r]{Targeted\\bridges} & \shortstack[r]{Mixed/\\audited} \\
\midrule
PDMX & 95/3/2 & 17/2/1 & 2/5 \\
LMD-full & 91/7/2 & 15/4/1 & 3/5 \\
Aria-MIDI & 93/4/3 & 16/3/1 & 2/5 \\
GigaMIDI & 94/5/1 & 16/4/0 & 2/5 \\
\bottomrule
\end{tabular*}
\end{table}

\newrev{Two reviewers reconciled assessments of 100 random edges, 20 targeted weak bridges, and five size-selected components per corpus (Table~\ref{tab:manual_audit_key}). Targeted samples describe failure modes, not prevalence; missed-version rates in unlabeled corpora remain unknown. Retrieval coverage, component membership, parse counts, and runtime are reported in \href{https://kunbai05-nb.github.io/}{Extended Results}. The 4,352 parse failures remain unresolved singletons.}

\newrev{Development-selected thresholds are frozen on held-out sources. No tested core setting meets the 0\% target. LMD-clean 1\%/5\% targets select thresholds 0.5/0.605, giving exposure 12.95\%/20.55\% and retention 92.63\%/99.24\%. At the 5\% target, ATEPP/ASAP/MAESTRO exposure is 0.36\%/0\%/0\% and retention 45.74\%/44.53\%/59.89\%. Thus development targets need not transfer. We distribute summary results; raw files remain subject to upstream licenses and third-party rights.}

\subsection{Downstream and Budget-Matched Cohorts}
\label{app:downstream_details}
\label{app:matched_reference}
\newrev{Table~\ref{tab:downstream_pools} gives the original five downstream pools; Fig.~\ref{fig:downstream_deviation} additionally includes calibrated union. Each condition/model/dataset uses three paired seeds and fixed evaluation content. Only random removal and MVA-SWI match file counts.}
\begin{table}[!tb]
\centering
\caption{\newrev{Training pools for the original five conditions. Exposure counts test works with a training version; retention is clean-token retention (\%). Union is the default embedding union.}}
\label{tab:downstream_pools}
\color{revisionblue}
\small
\setlength{\tabcolsep}{2.5pt}
\begin{tabular*}{\columnwidth}{@{\extracolsep{\fill}}lrrr@{}}
\toprule
Condition & Files & Exposure & Retention \\
\midrule
\multicolumn{4}{c}{ATEPP} \\
Uncorrected & 1223 & 121/121 & 100.00 \\
Random removal & 486 & 72/121 & 63.97 \\
MVA-SWI & 486 & 0/121 & 98.91 \\
Union & 209 & 18/121 & 30.88 \\
Label reference & 495 & 0/121 & 100.00 \\
\midrule
\multicolumn{4}{c}{LMD-clean} \\
Uncorrected & 1223 & 73/142 & 100.00 \\
Random removal & 1023 & 63/142 & 85.19 \\
MVA-SWI & 1023 & 25/142 & 99.67 \\
Union & 1107 & 50/142 & 99.53 \\
Label reference & 965 & 0/142 & 100.00 \\
\bottomrule
\end{tabular*}
\end{table}

\newrev{The separate ATEPP study (Table~\ref{tab:matched_reference_key}) matches MIDI-GPT initial weights across three seeds, 20k updates, 11,316 windows, and 5,884,369 scored tokens. The independent reference excludes validation/test works using labels and selects windows by deterministic hash under fixed quotas. The original reference repairs the MVA-SWI pool at equal file budget. Evaluation fixes 121 works, 152 files, 1,192 windows, and 602,471 tokens. Average NLL over valid tokens within files, files within works, then seeds. Paired work-bootstrap CIs condition on trained seeds.}
\begin{table}[!tb]
\centering
\caption{\newrev{Budget-matched ATEPP results. Dev. is mean signed per-work NLL deviation from the independent reference. Paired 95\% CIs condition on trained seeds.}}
\label{tab:matched_reference_key}
\color{revisionblue}
\small
\setlength{\tabcolsep}{2.5pt}
\begin{tabular*}{\columnwidth}{@{\extracolsep{\fill}}lrrr@{}}
\toprule
Condition & \shortstack[r]{Exposed\\works} & \shortstack[r]{Dev.\\(\%)} & 95\% CI (\%) \\
\midrule
Uncorrected & 119/121 & $-21.77$ & $[-22.93,-20.59]$ \\
Exact dedup. & 119/121 & $-21.77$ & $[-22.93,-20.59]$ \\
Random removal & 72/121 & $-10.11$ & $[-11.31,-8.93]$ \\
MVA-SWI & 0/121 & $+2.15$ & $[+1.85,+2.48]$ \\
Original ref. & 0/121 & $+0.89$ & $[+0.63,+1.19]$ \\
Indep. ref. & 0/121 & $0.00$ & $[0.00,0.00]$ \\
\bottomrule
\end{tabular*}
\end{table}

\newrev{Figure~\ref{fig:downstream_deviation} uses 1,223 Uncorrected ATEPP files with 121/121 exposed works, versus 1,140 and 119/121 in Fig.~\ref{fig:matched_reference}. Different pools, budgets, and references distinguish the studies; their absolute/signed deviations are not interchangeable. The former reports paired-seed SD, the latter work-bootstrap CIs.}

\section{Listening and Generation Evaluation}
\label{app:subjective_protocol}
\subsection{\rev{Generator Training and Decoding}}
\label{app:generator_training}
\newrev{Models train from scratch with a 237-token REMI-style vocabulary and 1,024-token context: 16 positions per $4/4$ bar, 128 pitches, 64 clipped durations, 16 velocity bins, eight pitched instrument groups plus drums, and bar/sequence controls. LMD uses four-bar prompts and up to 16 bars total. Final checkpoints score valid continuation tokens under teacher forcing.}

\newrev{MIDI-GPT~\cite{daotwenty_midigpt} uses 6 blocks, width 512, 8 heads, and 19.68M parameters. The reference Transformer has 10 pre-normalized causal blocks, width 384, 6 heads, feed-forward width 1,536, dropout 0.1, rotary positions, and tied embeddings. LSTM uses two 512-unit layers, embeddings of width 256, and dropout 0.1; TCN uses 9 residual blocks, width 384, feed-forward width 1,536, kernel 5, and dilations $2^0$--$2^8$. VAE has 2 encoder/3 decoder layers, width 128, 4 heads, latent width 64, and KL weight 0.01. Diffusion uses latent/model widths 64/256 and 1,000 linear noise steps ($10^{-4}$--0.02). Settings are in \texttt{configs/}; Transformer-S/M/L architectures are in checkpoint \texttt{model\_config}.}

\newrev{Controlled LMD training uses AdamW ($\beta_1=0.9$, $\beta_2=0.95$), batch 32/microbatch 8, gradient clipping 1.0, 1,000 warmup steps, and cosine decay to 20k updates. MIDI-GPT/LSTM/reference Transformer use learning rate $2\times10^{-4}$ and decay 0.1; VAE/diffusion use $3\times10^{-4}$/$5\times10^{-4}$ and decay 0.01. Decoding uses 128 prompt tokens and 512 new tokens, temperature 1.0, and top-$p=0.95$.}

\newpage % Keep the listening heading with its text in the final right column.
\subsection{Blinded Ratings and Uncertainty}
\newrev{Listening covers five conditions, two models/datasets, and ten trained seeds, excluding calibrated union. Of 97 recruits, 13 incomplete respondents and four failing at least two attention checks are excluded.}

\newrev{The 80 retained listeners split equally by musical background. Each rates 12 five-clip groups, two blind repeats, and three attention checks. The 96 groups contain 480 excerpts: 24 per condition/dataset/model from 12 cleaning-test prompts. Each excerpt receives ten ratings, five per background group, totaling 4,800 formal evaluations; 800 repeat ratings are excluded. Anonymous labels are shuffled. Eight-second clips use 120 BPM/16 kHz rendering, headphones, and full playback in a quiet setting; replay is allowed.}

\newrev{Ratings cover overall musicality, melody, rhythm, and harmony (1--5). ``Unable to judge'' is missing: 18,347/19,200 responses are valid, with 222/24/607 missing melody/rhythm/harmony ratings and complete overall ratings. Means use available responses; contrasts use matched nonmissing responses. Average within seeds, weighting seeds equally. Use 10,000 crossed listener--work bootstrap replicates with trained seeds fixed. Overall musicality and MVA-SWI minus Uncorrected are primary; the other three overall contrasts use simultaneous 95\% CIs. Secondary dimensions are descriptive. All four overall contrasts include zero in each setting; no equivalence margin was prespecified.}

\subsection{Automatic Metrics}
\label{app:generation_metrics}
\newrev{Evaluation fixes 120 ATEPP works/119 LMD-clean groups, five conditions, two models, ten seeds, and ten generations per group/checkpoint: 239,000 excerpts. Table~\ref{tab:passage_key} pools generations within seeds and reports seed means/SDs. A match requires at least 16 notes over four beats, allowing at most two pitch and two IOI discrepancies under a frozen rule. Full melody/rhythm similarity and FMD-CLaMP1~\cite{retkowski2025fmd,wu2023clamp} results and definitions appear in \href{https://kunbai05-nb.github.io/}{Extended Results}.}
\begin{table}[!tb]
\centering
\caption{\newrev{Long-passage match rates (\%, mean $\pm$ SD across ten training seeds). These measure agreement with held-out continuations, not perceived quality.}}
\label{tab:passage_key}
\color{revisionblue}
\small
\setlength{\tabcolsep}{2.5pt}
\begin{tabular*}{\columnwidth}{@{\extracolsep{\fill}}lrr@{}}
\toprule
Condition & MIDI-GPT & Transformer \\
\midrule
\multicolumn{3}{@{}l}{\textbf{ATEPP}} \\
Uncorrected & $22.42\,{\scriptstyle\pm\,0.92}$ & $23.89\,{\scriptstyle\pm\,0.71}$ \\
Random removal & $10.92\,{\scriptstyle\pm\,0.94}$ & $11.69\,{\scriptstyle\pm\,0.42}$ \\
MVA-SWI & $3.64\,{\scriptstyle\pm\,0.27}$ & $4.06\,{\scriptstyle\pm\,0.46}$ \\
Default union & $2.28\,{\scriptstyle\pm\,0.10}$ & $3.03\,{\scriptstyle\pm\,0.29}$ \\
Label reference & $3.33\,{\scriptstyle\pm\,0.17}$ & $3.94\,{\scriptstyle\pm\,0.43}$ \\
\midrule
\multicolumn{3}{@{}l}{\textbf{LMD-clean}} \\
Uncorrected & $21.90\,{\scriptstyle\pm\,1.86}$ & $22.30\,{\scriptstyle\pm\,1.40}$ \\
Random removal & $10.48\,{\scriptstyle\pm\,1.37}$ & $11.01\,{\scriptstyle\pm\,0.22}$ \\
MVA-SWI & $2.46\,{\scriptstyle\pm\,0.56}$ & $3.33\,{\scriptstyle\pm\,0.78}$ \\
Default union & $1.29\,{\scriptstyle\pm\,0.26}$ & $2.13\,{\scriptstyle\pm\,0.70}$ \\
Label reference & $2.41\,{\scriptstyle\pm\,0.49}$ & $3.22\,{\scriptstyle\pm\,0.90}$ \\
\bottomrule
\end{tabular*}
\end{table}

\end{document}